\documentclass[journal]{IEEEtran}

\usepackage{amsmath,tensor}
\usepackage{graphicx,amssymb,lineno}
\usepackage{psfig,subfigure}
\usepackage{verbatim}
\usepackage{slashbox}
\usepackage{threeparttable}
\usepackage{url}

\usepackage[usenames]{color}
\usepackage{colortbl}
\usepackage[table]{xcolor}

\definecolor{mygray}{gray}{.9}

\begin{document}

\title{A Cloud Infrastructure Service Recommendation System for Optimizing Real-time QoS Provisioning Constraints}

\author{
Miranda Zhang, Rajiv Ranjan, Michael Menzel, Surya Nepal, Peter Strazdins and  Lizhe Wang, \IEEEmembership{Senior Member,~IEEE}
\thanks{Miranda Zhang  and Peter Strazdins  are  with the Australian National University. 
 Rajiv Ranjan and  Surya Nepal  are   with the CSIRO, Australia.  
 Michael Menzel is with Karlsruhe Institute of Technology, Germany. 
   Lizhe Wang  is with the School of Information Science and Engineering, Yanshan University, Qinhuangdao, China. 
   Corresponding author: Lizhe Wang; Email: Lizhe.Wang@gmail.com
 }
 }

\markboth{IEEE Systems Journal,
Vol. x, No. x, xxx 2014}{Shell \MakeLowercase{\textit{et al.}}: Bare Demo of
IEEEtran.cls for Journals}

\maketitle

\begin{abstract}
Proliferation of cloud computing has revolutionized hosting and delivery of Internet-based application services. However, with the constant launch of new cloud services and capabilities almost every month by both big (e.g., Amazon Web Service, Microsoft Azure) and small companies (e.g. Rackspace, Ninefold), decision makers (e.g. application developers, CIOs) are likely to be overwhelmed by choices available. The decision making problem is further complicated due to heterogeneous service configurations and application provisioning Quality of Service (QoS) constraints. To address this hard challenge, in our previous work we developed a semi-automated, extensible, and ontology-based approach to infrastructure service discovery and selection based on only design time constraints (e.g., renting cost, datacentre location, service feature, etc.). In this paper, we extend our approach to include the real-time (run-time) QoS (end-to-end message latency, end-to-end message throughput) in the decision making process. Hosting of next generation applications in domain of on-line interactive gaming, large scale sensor analytics, and real-time mobile applications on cloud services necessitates optimization of such real-time QoS constraints for meeting Service Level Agreements (SLAs).  To this end, we present a real-time QoS aware multi-criteria decision making technique that builds over well known Analytics Hierarchy Process (AHP) method. The proposed technique is applicable to selecting Infrastructure as a Service (IaaS) cloud offers, and it allows users to define multiple design-time and real-time QoS constraints or requirements. These requirements are then matched against our knowledge base to compute possible best fit combinations of cloud services at IaaS layer.  We conducted extensive experiments to prove the feasibility of our approach.
\end{abstract}


\IEEEpeerreviewmaketitle

\thispagestyle{empty}

\section{Introduction}

In the cloud computing model, users access services according to their requirements, without the need to know where the services are hosted or how they are delivered. Increasing number of IT vendors (Amazon, GoGrid and Rackspace) are promising to offer applications, storage and computation resources as cloud hosting services. As a result, a large number of competing services are available for users \cite{ref1} to choose from. Naturally, it is challenging for users to select the right services that meet their QoS requirements in the service cycle from selection, deployment to orchestration (e.g. determine optimal web service when making service selection, identify suitable virtual machine servers for deploying web service instances, etc.) \cite{zhang2012investigating} . Effective service recommendation techniques are becoming important to help users (including developers) in their decision-making processes for critical application developments and deployments \cite{zhang2013investigating}. Such applications can include interactive games, real-time social networks, data analytics, scientific computing, business, Internet of Things (IoT) and other mobile applications as discussed next. All these applications have different needs and requirements.
\subsection{Motivation}
We next provide a few examples to demonstrate different types of applications with the needs to cater for real-time QoS requirements during their deployment lifecycle.

\textbf{Interactive Online Games:} In the gaming industry, World of Warcraft counts over six million unique players on daily basis. The operating infrastructure of this Massively Multiplayer Online Role Playing Game (MMORPG) comprises more than 10,000 computers \cite{nae2011dynamic}. Depending on the game, typical response times to ensure fluent play must remain below 100 milliseconds in online First Person Shooter (FPS) action games    \cite{beigbeder2004effects} and below 1-2 seconds for Role-Playing Games (RPGs). A good game experience is critical for keeping the players engaged, and has an immediate consequence on the earnings and popularity of the game operators. Failing to deliver timely simulation updates leads to a degraded game experience and triggers player departure and account closures    \cite{shaikh2006demand}. Startup gaming company with no existing infrastructure could launch a new game using public cloud infrastructure as cloud services offers the flexibility to scale on demand with no upfront investment. Using cloud services, the game application services can be dynamically allocated or de-allocated according to demand fluctuations. Game companies can also better serve the diverse international users with the global presence of data centers owned by Cloud providers.

\textbf{Real-time Mobile applications:} There is an explosion of (primarily mobile based) communication apps. For example, WhatsApp, acquired by Facebook, has  450 million users \cite{ref7}), Viber, acquired by Rakuten, has 200 million users \cite{ref8}) and WeChat, a Chinese rival, has 270 million users \cite{ref9}. For these apps, low latency (a QoS constraint) is very important for the real time collaboration experience. For example, video conferencing, has a limit of about 200 to 250 milliseconds delay for a conversation to appear natural \cite{weinman2011time}. These apps have similar requirements as the game apps. They require large number of servers to support millions of users, need optimization on latency, speed and throughput. It's worth mentioning that even for a generic web application, there are experiments with delaying the page in increments of 100 milliseconds and found that even very small delays would result in substantial and costly drops in revenue    \cite{weinman2011time}.

\textbf{Big Data, IoT (Internet of Things) and eScience:} We are closing in on the transfer of a zettabyte of data annually \cite{ref11}, resulting from internet search, social media, business transactions, and content distribution. Similarly, scientific disciplines increasingly produce, process, and visualize data sets gathered from sensors \cite{GHadoop}. If the prediction holds true, then the Square Kilometer Array (SKA) radio telescopes will transmit 400,000 petabytes ($\sim$400 exabytes) per month or a whopping 155.7 terabytes per second \cite{ref12}. Furthmore, European Space Agency (ESA) will launch several satellites in the next few years \cite{ref13}, which will collect data about the environment, such as air temperatures and soil conditions, and stream that data back in real time for analysis.  Similarly in the finance industry, New York Stock Exchange creates 1 terabyte of market and reference data per day covering the use and exchange of financial instruments. On the other hand, Twitter feeds generate 8 terabytes of data per day of social interactions \cite{ref14}. Such ``Data Explosions'' has led to research issues such as: how to effectively and optimally manage and analyze such large amount of data. The issue is also known as the ¡®Big Data' problem    \cite{hey2003data}, which is defined as the practice of collecting complex data sets so large that it becomes difficult to analyze and interpret manually or using on-hand data management applications (e.g., Microsoft Excel). As both storing and analyzing the data requires massive amount of storage capacity and processing power. Companies and/or institutions may want to offload the complexity of managing hardware infrastructure to Cloud providers who are specialized in that, plus eliminating the need to wait for facilities to be built.

\textbf{Other:} Apart from the above mentioned scenarios, there are many more cases our proposed solution would be useful.

A stock investor, individual or firm, may want to test out a new strategy for monitoring analyzing data which automatically triggers alert when certain price pattern or keyword is identified in the source data. This may require a lot of compute resources periodically.
System administrators and developers may need a lot of simulated clients from all around the world for a website load testing before its official release.

A bitcoin    \cite{bedford2013bitcoin} (or some other similar cryptocurrencies \cite{ref17}) miner may decide to invest on some additional resource in mining when the price of the currency is high, and stop the mining when the profit does not justify the expense anymore.

\subsection{The Problem}
While the elastic nature of cloud services makes it suitable for provisioning aforementioned applications, the heterogeneity of cloud service configurations and their distributed nature raises some serious technical challenges. In particular, we deal with following research problems:

\textbf{Selecting Optimal Service Configuration: }The cloud computing landscape is evolving with multiple and diverse options for compute (also known as virtual machines) and storage services. Hence, application owners are facing a daunting task when trying to select cloud services that can meet their constraints. According to Burstorm \cite{ref18} there are over 426 of various compute and storage service providers with deployments in over 11,072 locations. Even within a particular provider there are different variations of the services. For example, Amazon Web Service (AWS) has 674 different offerings differentiated by price, QoS features and location \cite{ref1}. Add to this every quarter they add about 4 new services, change business models (price and terms) and sometimes even add new locations. To be able to select the best mix of service offering from an abundance of possibilities, application owners must simultaneously consider and optimize complex dependencies and heterogeneous sets of criteria (price, features, location, QoS etc.). For instance, it's not enough to just select optimal cloud storage service, corresponding computing capabilities are essential to guarantee that one is able to process the data as fast as possible while minimizing the cost.

\textbf{Incorporating Network QoS-awareness in Service Selection Process: }As the cloud data centers are distributed across the Internet, the network QoS (data transfer latency) varies. This variation is dependent upon the location of data center and location of input data stream. Current approaches do not differentiate between the QoS of compute and storage services and the QoS of the wide area network that interconnects input data stream sources to cloud data centers. This raises a research question: how to optimize the process of choosing the best compute and storage services, which are not only optimized in terms of price, availability, processing speed but also offers good QoS (e.g. network throughput and response delivery latency)?

\begin{table*}[!ht]
\caption{A BRIEF COMPARISON OF THE CLOUD RECOMMENDER WITH OTHER EXISTING SOLUTIONS}\label{comparison}
\begin{center}
\begin{tabular}{|l|l|l|l|l|}

\hline \backslashbox{Product}{Feature}  &  $\begin{array}{l}
{\rm{QoS}}\\
{\rm{Benchmark}}
\end{array}$   &
$\begin{array}{l}
{\rm{Single  Criteria }}\\
{\rm{Comparison}}
\end{array}$   & $\begin{array}{l}
{\rm{Aggregate  Ranking }}\\
{\rm{\&  Comparison}}
\end{array}$  & $\begin{array}{l}
{\rm{Cloud}}\\
{\rm{Management}}
\end{array}$  \\
\hline $\;$ Broker@Cloud & \multicolumn{4}{c|}{No evidence on progress of project}\\
\hline $\;$ Yuruware & No & No & No & \cellcolor{mygray}Yes\\

\hline $\;$ CloudHarmony & \cellcolor{mygray}Adjustable & No & No & No\\
\hline $\;$ Cloudorado & No & \cellcolor{mygray}Yes & No & No\\
\hline $\;$ CloudBroker & \cellcolor{mygray}Adjustable & \cellcolor{mygray}Yes & No & No\\
\hline $\;$ CloudRecommender & \cellcolor{mygray}Fixed & \cellcolor{mygray}Yes & \cellcolor{mygray}Yes & No \\
\hline
\end{tabular}
\end{center}
\label{t:1}
\end{table*}

\subsection{Our Contributions }
We propose a new technique that aids in network QoS-aware selection of cloud services for provisioning mobile (or device with internet access but limited processing capability and storage), real-time and interactive applications. We build upon our previous work \cite{zhang2013investigating} where we have developed an automated approach, along with a unified domain model capable of fully describing infrastructure services in Cloud computing    \cite{zhang2012ontology}   \cite{zhang2012declarative}. While our previous approach supports simple cloud infrastructure service selection based on declarative Structured Query Language (SQL), it does not take into account real-time, variable network QoS constraints. Furthermore, a declarative SQL-based selection approach only allows users to compare and select a cloud service based on a single criterion (e.g. total cost, max size limit for storage, memory size for compute instance). In other words, our previous approach was not capable of supporting a utility function that combines multiple selection criteria pertaining to storage, compute, and network services.  In this paper, we make following concrete contributions:

\textbf{1. Problem Formulation.} We provide a clear formulation of the research problem by identifying the most important cloud service selection criteria relevant to specific real-time QoS-driven applications, selection objectives, and cloud service alternatives.

\textbf{2. Multi-criteria QoS Optimization.} We adopt and implement an Analytic Hierarchy Process (AHP) based decision (service selection) making technique that handles multiple quantitative (i.e. numeric) as well as qualitative (descriptive, non numeric, like location, CPU architecture: 32 or 64 bit, operating system) QoS criteria. AHP determines the relative importance of criteria to each user by conducting pair-wise comparisons.

\textbf{3. Network-aware QoS Computation.} We implement a generic service that helps in collecting network QoS values from different points on the Internet (modeling big data source location) to the cloud data centers.

The paper is structured as follows. In section \ref{related}, we survey the state-of-the-art in Cloud Service Selection and Comparison (CSSC) techniques. We also highlight their significant limitations, their relationship and dependency on some of the prior concepts from other fields in computing. In Section \ref{model}, we present the extension we made to our previously proposed decision making framework. We also explain the benefits of applying AHP and importance of considering QoS. In section \ref{experiment}, we present evaluations (conducted in real-world context) of the proposed decision support tool and techniques, which will automate and map users' specified application requirements to specific Cloud service configurations. In section \ref{conclusion}, we conclude and point out open research questions and future directions in this increasingly important area.


\section{BACKGROUND AND RELATED WORK}\label{related}

Though branded calculators are available from individual cloud providers, such as Amazon \cite{ref21} and Azure \cite{ref22} for calculating service leasing cost, it is not easy for users to generalize their requirements to fit different service offers (with various quota and limitations), let alone computing and comparing costs. A number of research    \cite{li2010cloudcmp} and commercial projects (mostly in their early stages) provide simple cost calculation or benchmarking and status monitoring, but none is capable to consolidate all aspects and provide a comprehensive ranking of infrastructure services. For instance, CloudHarmony \cite{ref24} provides up-to-date benchmark results without considering cost, Cloudorado \cite{ref25} calculates the price of IaaS-level CPU services based on static features (e.g., processor type, processor speed, I/O capacity, etc.) while ignoring dynamic QoS features (e.g. latency, throughput etc.). Yuruware \cite{ref26} used to provide a Compare service during beta version in 2012 (now removed or integrated into another service). Although they aim to provide an integrated tool with monitoring and deploying capabilities, it is still under development. One other similar system is Swinburne University's Smart Cloud Broker Service \cite{ref27}, from the screencast they released, we can tell that their benchmarking is done in real-time which means users have to wait for the results to come back. We have considered this kind of situations, but decided to collect the benchmarking result beforehand. Because this way no matter how many cloud providers users want to compare against, they can still get the result with minimum (or no) waiting time. Another reason we choose to do it this way is because, at any particular point in time, the network benchmark result is not conclusive as performance fluctuates during time, so we use aggregated average which is a more reliable overall indication.

To further distinguish ourselves from others, we offer the following two innovative features when ranking, selecting, and comparing various vendor services: 1) allow users to choose to include the QoS requirements during comparison; 2) when users want to take into account mixed qualitative (e.g. hosting region, operating system type) and quantitative criteria, we apply the Analytic Hierarchy Process (AHP) to aggregate numerical measurements and non numerical evaluation. Results are personalized according to each user's preferences, because AHP takes users' perceived relative importance of criteria (pair-wise comparisons) as inputs.

Table \ref{comparison} shows a brief comparison of the CloudRecommender with other existing products we mentioned previously. We have to clarify that we are more interested in the first 3 features. Yuruware had claimed to have comparison features in the past, but removed later.

Menzel and Ranjan    \cite{menzel2012cloudgenius} introduced a framework called ``CloudGenius'' that supports decision making process on web server migration into the cloud. Our system supplements and partially extends their work. While ``CloudGenius'' focus on Virtual Machine (VM) selection, means it considers the software requirements (i.e. operating system version, supported languages), our study focus more on the hardware requirements (i.e. size of memory and hard disk).  Although we have borrowed the idea of using the AHP (with simplification) for rank calculation from ``CloudGenius'', we used it differently, as we applied the method in our declarative program which mainly handles data and calculation with database and SQL. That means it may be easier to scale out the solution using Hive \cite{ref29} with minimal change, as suppose to rewrite the java code to fit the Map Reduce Framework    \cite{dean2008mapreduce}.

Queuing theory is one of the much studied method in QoS modeling and control from the infrastructure system administrator perspective    \cite{sha2002queueing} but our case is different, because we have no control of the infrastructure. Since we can only measure the QoS, we collected the statistics using the ``speedtest"  service provided by CloudHarmony due to easy adoption and ever evolving nature of this service. Klein et al.    \cite{klein2012towards} proposed a highly theoretical model based on Euclidean distance for estimating latency, which we believe have omitted too much details to be practically accurate. However, we can use this model to estimate latency when QoS data is not available for a new client location.

There are methods proposed for network aware service composition    \cite{yu2007efficient}   \cite{zeng2004qos}    \cite{zheng2013qos} considering generic web service, i.e. at the Software-as-a-Service (SaaS) and Platform-as-a-Service (PaaS) level. But the compatibility constrains at the IaaS level are different from web service. For example, generic web services are distinguished by their features, QoS and prices. It does not make sense to include 2 exact same services in one composition as one job does not need to be done twice, but using multiple quantity of an IaaS offer is perfectly valid.

\begin{table}[!h]
\begin{center}\caption{SYMBOLS USED IN THE FORMULAS} \label{table:formula_symbols}
\begin{tabular}{|c|p{6cm}|}
\hline
\textbf{Symbol }&  \textbf{Meaning  }  \\
\hline a & Resource usage behave like a decision variable. \\
\hline C & Set of all possible Cloud providers. \\
\hline c & Cloud Provider, e.g. Amazon,Rackspace, GoGrid. \\
\hline D & Downloading speed. \\
\hline i & Identifies a request. \\
\hline L & Set of all possible datacenter locations. \\
\hline l & A datacenter location, e.g. Sydney, Tokyo. \\
\hline $\zeta$ & Latency (download). \\
\hline  M & Memory Size (e.g. 8G). \\
\hline  P & Price \\
\hline  R & Set of all possible resources, including all types whether it is Compute, Storage or Network.\\
\hline  r & Identifies a source, e.g. GoGrid XX - Large Instance, S3 Storage Serive, EC2 instance.\\
\hline $\gamma $ & $\;$ Set of Requests from one user. \\
\hline  S & Storage. \\
\hline  T & Period of time the resource is used. \\
\hline  t & Exact point in time, like a time stamp. \\
\hline  U & CPU speed. \\
\hline $\mu$ & Uploading speed. \\
\hline w & Weight. \\
\hline
\end{tabular}
\end{center}
\end{table}

\section{System Design}\label{model}

This section will describe our system's architecture and give details on how it's realised, i.e. formulas on how weight, rating, cost are calculated. We keep all the formulas in subsection \ref{subsec:formal_model}, then we show where/in which step different formulas are applied and how relates to each other in subsection \ref{subsec:algorithm}. In the last subsection, we provide illustrations of overall system design and include any worth mentioning details that does not fit into the previous subsections.

\subsection{Formal Model }
\label{subsec:formal_model}
To give a conceptual explanation of our approach to address the QoS optimization problem, we define a formal model in this section. Based on the formal model, we can describe the involved concepts that are incorporated in the algorithm presented later. Particularly, we define a cost estimation function using resource utilization estimations, and a benefit-cost ratio-based evaluation function which considers weights. Furthermore, we present a pair-wise comparison method to calculate normalized weights. For more precise resource utilization estimations, we show how variable resource utilization patterns can be incorporated into cost estimation.

\subsubsection{Cost Estimation}

Let ``a" be the resource usage of a particular resource from a data center location of a Cloud provider. For example, we can use $a_{storage,any,any}=50 GB$ to represent user's need to store 50 GB of data in the cloud. The symbols' meanings are summarized in Table \ref{table:formula_symbols}. Equation \ref{eq:a} means the usage of the compute resource $r$ from provider $c$ at location $l$ is between $0$ and $n$. This value is usually suggested by users. Our assumption is that users may have a rough estimate of how much resources they might need.
\begin{equation}\label{eq:a}
{{\rm{a}}_{r,c,l}} \in \{ 0,1, \ldots ,n\}
\end{equation}
To calculate the Cost (represented by function: $\wp $) for one kind of resource used at one point in time, we multiply its usage with the corresponding unit price (P) as:
\begin{equation}\label{eq:p}
\wp (t) = {a_{r,c,l}}{P_{r,c,l}}
\end{equation}

After initial filtering on which options are appropriate for users, we can calculate the total (minimum) price per unit time for desired resource(s) (assume constant resource usage pattern throughout the time) as in formula \ref{eq:period_cost}. We assume users will choose the time period (T) they want to estimate price for, e.g. 1 hour, 30 days.
\begin{equation}\label{eq:period_cost}
{a_{r,c,l}}{P_{r,c,l}}{T_{r,c,l}}
\end{equation}

\begin{figure}[!h]
 \centering
 \includegraphics[scale=0.5]{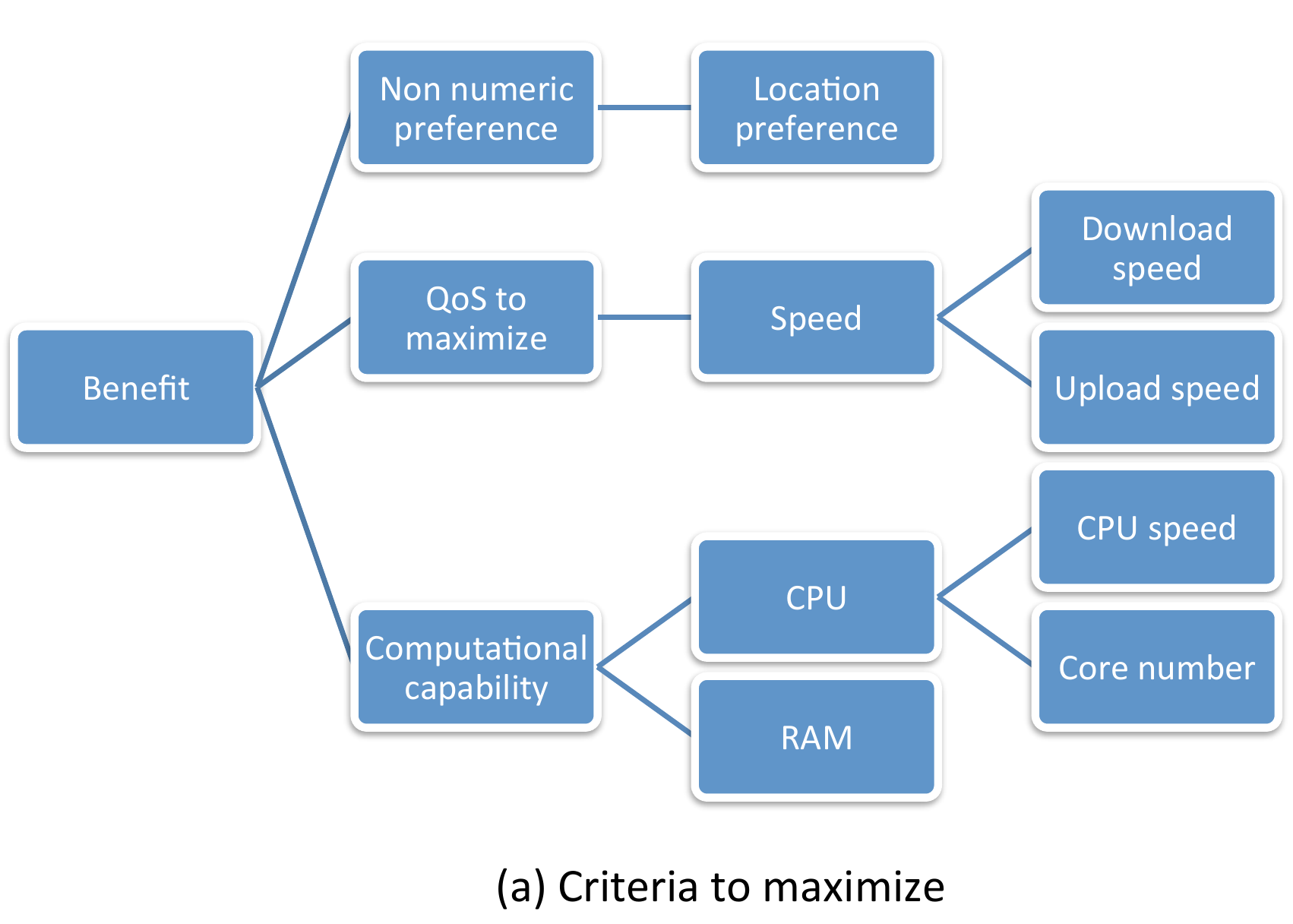}
  \includegraphics[scale=0.5]{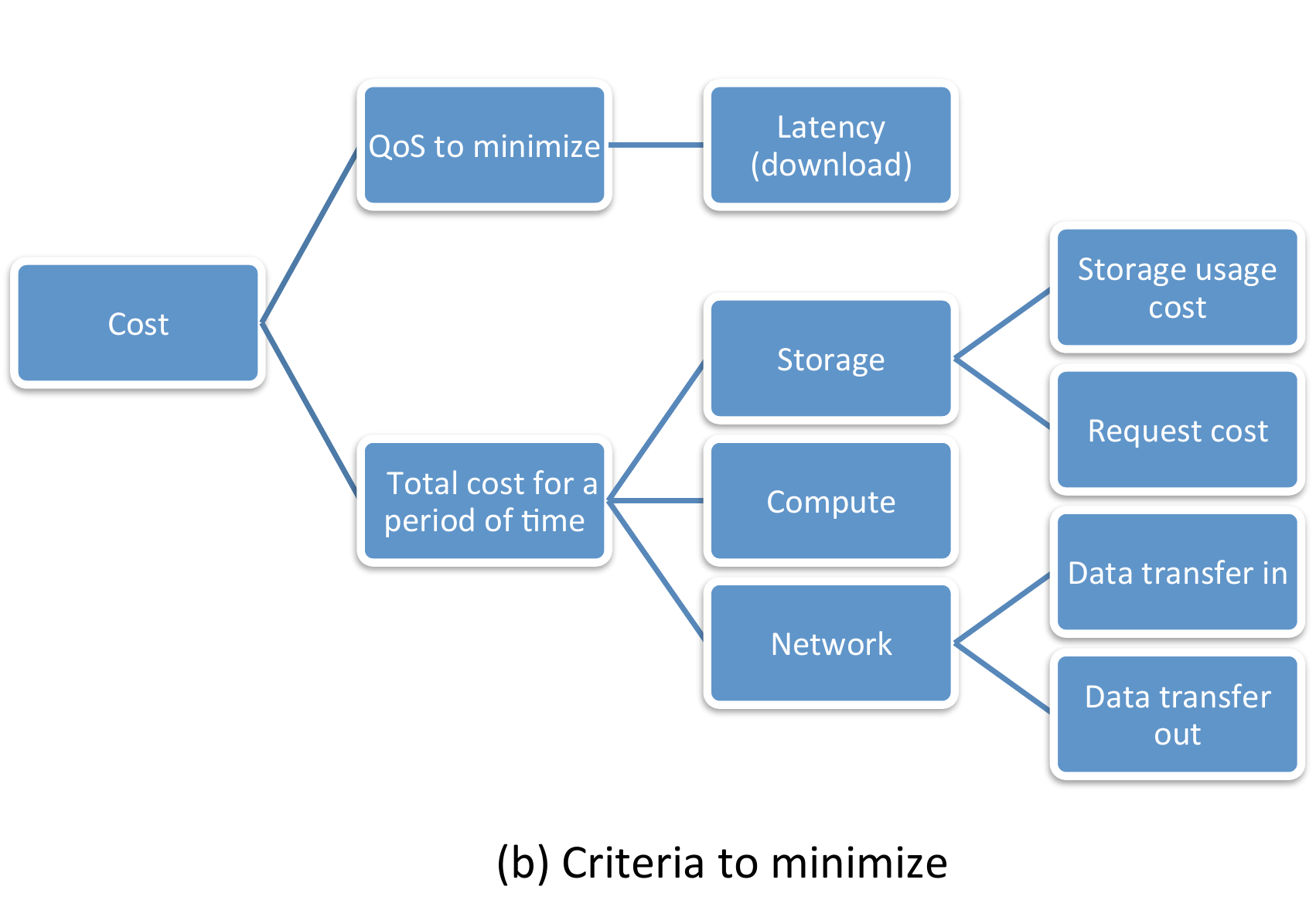}
 \caption{ Criteria taken into consideration during comparison. There are 2 categories: benefit and cost. ``Benefit" groups the ``good" criteria which are meant to be maximized. Similarly, ``Cost" groups the ``bad" criteria to be minimized. The actual values to be collected and stored are at the ``leaf" (i.e. Node/criterion with no children) of the ``tree". For example, under ``Benefit", numeric values are collected for ``Download/Upload Speed", ``CPU Speed" and ``Number of Cores". ``QoS to Maximize" is the parent/big category ``Download/Upload Speed" belongs to, there is no value stored for this node.}
\label{fig4}
\end{figure}

\subsubsection{Cost Benefit Ratio}
\label{subsec:cost_benefit_ratio}

In our decision making framework, we consider the following QoS statistics: download latency ($\zeta$), download speed (D) and upload speed ($\mu$). Those characteristics are important for end-users experience and satisfaction. It's possible to have options that have small price difference, or when having high quality service is more important than saving money. So we offer to calculate the cost/benefit ratio for the resources requested as in equation \ref{eq:ratio}.
\begin{equation}\label{eq:ratio}
\frac{ w_1 \sum { a_{c,l,r} P_{c,l,r} T_{c,l,r} } + w_2 \bar\zeta_{c,l,r} } {w_3 \bar\mu_{c,l,r} + w_4\bar D_{c,l,r}}
\end{equation}

Since users are likely to select a combination of compute storage and network services, hence the summation over resources when calculating the cost.

Note that the network QoS of Compute and Storage Service are both collected then separately stored, since user maybe only interested in one of the services. For example, transferring files from (and to) the compute instance relatively ``local"  mounted storage is different from downloading or uploading files from/to dedicated storage only service (like AWS S3 \cite{S3}). In case user select both, we use the average. For instance, in the equation we used $\bar D$ to denote that we take the average of $D_{compute}$(download speed measured from the Compute service) and $D_{storage}$  (download speed measured from the Storage service).

Symbol w represents the weight, which measures users' perceived importance on a parameter, and $w_{1}+ w_{2}=1$ and $w_3+ w_4=1$ means the sum of the weights of benefits and cost each equals to one. Fig. \ref{fig4} shows the criteria to be optimized. They are categorized into two groups: to be maximized or to be minimized.

As we named this ratio ``Cost Benefit Ratio'', we put cost on the numerator and benefit in the denominator. As a result we will be looking for smaller ratio as better option. Reversing numerator and denominator can still work, just means bigger ratios indicating better option.

\begin{table}[!h]
\begin{center}\caption{ABSOLUTE VALUE AND CORRESPONDING DESCRIPTIVE SCALE REPRESENTING RELATIVE IMPORTANCE} \label{table:scale}
\begin{tabular}{|c|c|c|}
\hline
\textbf{Scale }&  \textbf{Value } & Reciprocals\tnote{1}$*$ \\
\hline equal & 1 & 1\\
\hline moderate & 3 & 1/3 \\
\hline strong & 5 & 1/5 \\
\hline very strong & 7 & 1/7 \\
\hline extreme & 9 & 1/9\\
\hline
\end{tabular}
\begin{tablenotes}
  \item[1] $*$If activity i has one of the above nonzero numbers assigned to it when compared with activity j, then j has the reciprocal value when compared with i.
\end{tablenotes}

\end{center}
\end{table}

\subsubsection{Weight computed by Pairwise Comparison}
\label{subsec:weight&pairwise_comparison}

The weight is calculated based on AHP's pair wise comparison method. We choose the commonly used scale    \cite{ghodsypour1998decision}   \cite{haas2005illustrated} shown in Table \ref{table:scale}. In case user chooses to treat all options equally, (4) become (5).
\begin{equation}
\frac{ 0.5\sum {{a_{r,c,l}}{P_{r,c,l}}{T_{r,c,l}}} + 0.5{\zeta_{c,l,r} }}{{ 0.5{{\bar \mu }_{c,l,r}} + 0.5{{\bar D}_{c,l,r}}}}
\end{equation}

\begin{table}[!h]
\begin{center}\caption{SYMBOLS USED IN WEIGHT EXPLANATION } \label{table:weight_explanation_symbols}
\begin{tabular}{|c|c|c|}
\hline
\textbf{Symbol }&  \textbf{Meaning } \\
\hline $\tau $ & $\sum\limits_{n = 1}^{n = 4} {{y_n}} $ \\
\hline V & Value given by User to rate the importance. \\
\hline $V_{compute_{disk}}$ & How important is the size of disk space on VM. \\
\hline $V_{cost}$ & Importance value for cost. \\
\hline ${V_{latency}}$ & Importance value for Download Latency. \\
\hline $V_{ram}$ & How important is the size of memory allocated to VM. \\
\hline $V_{speed_{upload}}$ & Importance value for Upload Speed. \\
\hline $V_{speed_{download}}$ & Importance value for Download Speed. \\
\hline x & Some user input value. \\
\hline y & Sum of the row values. \\
\hline ${y_1}$ & $\left( {\sum\limits_{n = 1}^{n = 3} {{x_n}} } \right) + 1$ \\
\hline ${y_2}$ & $\left( {\sum\limits_{n = 4}^{n = 5} {{x_n}} } \right) + 1 + \frac{1}{{{x_1}}}$\\
\hline
\end{tabular}
\end{center}
\end{table}

\begin{table*}
\begin{center}
\caption{MATRIX ILLUSTRATING HOW TO TURN PAIR-WISE PREFERENCE INTO GLOBAL WEIGHT}
\label{table:weight}
\begin{tabular}{ccccccc}
& $V_{speed_{upload}}$ & $V_{speed_{download}}$ & $V_{ram}$ & $V_{compute_{disk}}$ & Row Sum & Weight \\
$V_{speed_{upload}}$ & 1 & ${x_{1}}$ & ${x_{2}}$ & ${x_{3}}$ & ${y_{1}}$ & ${y_1}/\tau$  \\
$V_{speed_{download}}$ & $1/{x_1}$ & 1  & ${x_{4}}$ & ${x_{5}}$ & ${y_{2}}$ & ${y_2}/\tau$  \\
$V_{ram}$ & $1/{x_2}$ & $1/{x_4}$ & 1 & ${x_{6}}$ & ${y_{3}}$ & ${y_3}/\tau$ \\
$V_{compute_{disk}}$ & $1/{x_3}$ & $1/{x_5}$ & $1/{x_{6}}$ & 1 & ${y_{4}}$ & ${y_4}/\tau$ \\
& & & & Column Sum & $\tau$ & 1\\
\end{tabular}
\end{center}
\end{table*}

Otherwise, weight is calculated as shown in Table \ref{table:weight} on page \pageref{table:weight}. 
The meaning of symbols is explained in Table \ref{table:weight_explanation_symbols}.
 
The fully fledged AHP method consists of repeated matrix squaring to compute the eigenvector, see \ref{eq:eigenvector}, every time the eigenvector gain a tiny improvement on precision at the cost of expensive computation, this is supposed to be repeated until no big enough difference (i.e. to four decimal places)  can be observed. In our case, we noticed that the improvement is so small that this rule can be relaxed to omit iterations on matrix squaring.

\begin{equation}\label{eq:eigenvector}
\left[ \begin{array}{l}
{y_1}/\tau \\
{y_2}/\tau \\
{y_3}/\tau \\
{y_4}/\tau \\
\end{array} \right]
\end{equation}

\begin{table}
\caption{EXAMPLE USER PREFERENCE}
\label{table:weight_example}
\begin{tabular}{@{}c@{}c@{}c@{}c@{}c@{}}
& $V_{speed_{upload}}$ & $V_{speed_{download}}$ & $V_{ram}$ & $V_{compute_{disk}}$ \\
$V_{speed_{upload}}$ & 1 & 1/3 & 1/5 & 1/5 \\
$V_{speed_{download}}$ & & 1 & 3 & 5 \\
$V_{ram}$ & & & 1 & 3 \\
$V_{compute_{disk}}$ & & & & 1 \\
\end{tabular}
\end{table}

For example, user may have preference like shown in Table \ref{table:weight_example}. It will produce the preference matrix $M_1$.

\begin{equation}\label{matrix:m1}
\begin{array}{c}
    M_1\\
    \begin{bmatrix}
        1 & 1/3 & 1/5 & 1/5 \\
        3 & 1   & 3   & 5 \\
        5 & 1/3 & 1   & 3 \\
        5 & 1/5 & 1/3 & 1\\
    \end{bmatrix}
\end{array}
\end{equation}

Table \ref{table:example_eigenvector} shows the steps breakdown to compute the eigenvector from \ref{matrix:m1} before matrix squaring.

\begin{table}[h]
\caption{EXAMPLE EIGENVECTOR CALCULATION}
\label{table:example_eigenvector}
\begin{tabular}{cccccccccc}
          &   &        &   &        &   &     &            & Row Sum \\
        1 & + & 0.3333 & + & 0.2    & + & 0.2 & =          & 1.7333  \\
        3 & + & 1      & + & 3      & + & 5   & =          & 13      \\
        5 & + & 0.3333 & + & 1      & + & 3   & =          & 9.3333  \\
        5 & + & 0.2    & + & 0.3333 & + & 1   & =          & 6.5333  \\
          &   &        &   &        &   &     & Column Sum & 30.5999 \\        
\end{tabular}
\end{table}

The result eigenvector would be:

\begin{equation}
\label{eigenvector1}
v_1=
\left[ 
\begin{array}{c} 
0.0566 \\ 
0.4248 \\
0.3050 \\
0.2135
\end{array} 
\right] 
\end{equation}

If we square the matrix $M_1$ we get:

\begin{equation}
\label{eq:m1square}
\begin{array}{c}
    M_2\\
    \begin{bmatrix}
        4      & 58/75 & 22/15  & 8/3 \\
        46     & 4     & 124/15 & 98/5 \\
        26     & 44/15 & 4      & 26/3 \\
        184/15 & 98/45 & 34/15  & 4\\
    \end{bmatrix}\\
    M_1 \times M_1
\end{array}
\end{equation}

The eigenvector calculated from $M_2$ is:

\begin{equation}
\label{eigenvector2}
\left[ 
\begin{array}{c} 
0.0597 \\ 
0.5223 \\
0.279 \\
0.1389
\end{array} 
\right] 
\end{equation}

The change of value in the new eigenvector is vary small, hence why we decide to omit this step and just use the original weight values ($v_1$). And we assume the preference for cost and latency are 0.8 and 0.2, so we can calculate the overall rank as shown in equation \ref{eq:ratio_example} :

\begin{multline}\label{eq:ratio_example}
( 0.8 \sum { a_{c,l,r} P_{c,l,r} T_{c,l,r} } + 0.2 \bar\zeta_{c,l,r} ) \\ 
( 0.0566 \bar\mu_{c,l,r} + 0.4248\bar D_{c,l,r} \\ 
+ 0.3050 \sum M_{c,l,r} + 0.2135 \sum S_{c,l,r} )^{-1} 
\end{multline}

Where $M$ represents memory size and $S$ is storage size.

\begin{table}[!ht]
\begin{center}\caption{SYMBOLS USED IN ALGORITHM  } \label{table:algo_symbols}
\begin{tabular}{|c|p{6cm}|}
\hline
\textbf{Symbol }&  \textbf{Meaning} \\
\hline AvgQoS &   Table/Relation contains the QoS data collected \\
\hline ${D_{compute}}$ &   Download speed from the compute instance \\
\hline ${D_{storage}}$ &   Download speed from pure storage, i.e. S3 \\
\hline $\bar D$ &   Average download speed calculated as: $\frac{1}{2}\left( {{D_{{\rm{compute}}}} + {D_{storage}}} \right)$ \\
\hline $\ell$ &  
$\ell  \subseteq L$  Some set of locations which are specified by the user, by default 
 $\ell  = L$, which means consider all locations available.
 \\
\hline $M_{min}$ & Minimum memory requirements of compute instance/server  \\
\hline $pric{e_{\max }}$ &   The maximum price one is willing to spend \\
\hline $\rho$ &   
$\rho  \subseteq C$ Some set of  Cloud  providers which are specified by  the user; by
default   $\rho  \subseteq C$, which means consider all locations available. \\
\hline ${\Re _{compute}}$ &  Table/Relation contains all data collected about Compute resources.  \\
\hline ${\Re _{network}}$ &  Table/Relation contains all data collected about Network resource. \\
\hline ${\Re _{storage}}$ & Table/Relation contains all data collected about Storage resource. \\
\hline $\bar \mu $ &   Average upload speed, similar to $\bar D$ \\
\hline U & A tuple representing the estimated usages provided by user, containing the
following: $( U_{compute} , U_{storage} , U_{data_{in}} , U_{data_{out}} )$\\
\hline W & A tuple representing the preference/weight given to each component by the user, it consists of the following: 
$( W_{compute} , W_{storage}, W_{network} $ , $ W_{download} , W_{upload}, W_{latency} )$ \\
\hline
\end{tabular}
\end{center}
\end{table}

\begin{table}[!h]
\begin{center}\caption{SYMBOLS USED IN ALGORITHM: RELATIONAL ALGEBRA AND SET OPERATIONS} \label{table:relational_algebra_set}
\begin{tabular}{|c|p{6cm}|}
\hline
\textbf{Symbol }& \textbf{Meaning } \\
\hline 
G & 
    Aggregation operation over a schema, like a \textbf{group by} clause in SQL.
    It follows the format:
    $\tensor[_{a_g}]{G}{_{agg\_op(attri)}} (r)$
    where $a_g$ is the grouping attribute.
    $agg\_op(attri)$ is the aggregation operation over attribute (attri). 
    There are five aggregate functions that are included with most relational database systems.
    These operations are Sum, Count, Average, Maximum and Minimum.
    \textbf{r} is an arbitrary relation.
    See relationa algebra wiki page \cite{ref36} for more details.\\
    
\hline $\sigma $ &  Selection, see \cite{ref36}.\\
\hline $\bowtie$ & Natural join: depends on the condition can be either $\theta$-join or equijoin. For example, $\bowtie(Provider,Location)$ means equijoin where the condition is join only under the same provider and location\\
\hline  $ \cup$  & Set union operation. \\
\hline $\mapsto $ & Ordered pair, here we use it to denote a new record being formed.\\
\hline
\end{tabular}
\end{center}
\end{table}

\begin{table*}[!ht]
\begin{center}
\begin{tabular}{l l l l l}
\multicolumn{5}{l}{ Algorithm 1: orderedSolutions $( \ell , {M_{\min }} , price_{\max} , \rho , U , W )$ } \\
\hline  
1 & \multicolumn{4}{l}{//Filtering on the static characteristics}\\
2 & \multicolumn{4}{l}{${\Phi _{compute}}: = {\sigma _{provider \in \rho  \wedge location \in \ell  \wedge memory \ge {M_{\min }}}}\left( {{\Re _{compute}}} \right)$}\\

3 & \multicolumn{4}{l}{//Link it with QoS statistics.}\\
4 & \multicolumn{4}{l}{${\varphi _{compute}}: = {\Phi _{compute}}{{\rm \bowtie}_{provider,location,serviceName}}AvgQoS$}\\

5 & \multicolumn{4}{l}{${\Phi _{storage}}: = {\sigma _{provider \in \rho  \wedge location \in \ell  \wedge quot{a_{low}} < {\upsilon _{storage}}}}\left( {{\Re _{storage}}} \right)$}\\

6 & \multicolumn{4}{l}{//Calculating storage price for each tier.}\\
7 & \multicolumn{4}{l}{${\rm{storageCostByQuota: = emptylist}}$}\\
8 & \multicolumn{4}{l}{$foreach\;{\zeta_s} \in {\Phi _{storage}}\;do$}\\
9 & $|$ & \multicolumn{3}{l}{$if\;quot{a_{\min }}({\zeta_s}) < {U_{storage}}$}\\
10 & $|$ & $|$ & \multicolumn{2}{l}{$foreach\;{\zeta_s} \in {\Phi _{storage}}\;do$}\\
11 & $|$ & $|$ & $|$ & \multicolumn{1}{l}{$
storageCostByQuota: = storageCostByQuota \cup \left\{ {{\zeta_s} \mapsto quot{a_{\max }}({\zeta_s})} \right\}* unitPrice({\zeta_s})$}\\
12 & $|$ & $|$ & \multicolumn{2}{l}{else}\\
13 & $|$ & $|$ & $|$ & \multicolumn{1}{l}{$\begin{array}{l}
storageCostByQuota: = storageCostByQuota \cup \left\{ {{\zeta_s} \mapsto \left( {{U_{storage}} - quot{a_{\min }}({\zeta_s})} \right)} \right\}*unitPrice({\zeta_s})
\end{array}$}\\

14 & $|$ & $|$ & \multicolumn{2}{l}{end}\\
15 & $|$ & \multicolumn{2}{l}{end}\\
16 & \multicolumn{4}{l}{end}\\

17 & \multicolumn{4}{l}{//Combining storage cost in different tiers to get total.}\\
18 & \multicolumn{4}{l}{$\begin{array}{l}
storageCost: = \;\;\;\;\;\;\;\;\;\;\;\;\;\;\;\;\;\;\;\;\;\;\;\;\;\;\;\;\;\;\;\;\;\;\;\;\;\;\;\;G\;\;\;\;\;\;\;\;\;\;\;\;\;\;\;\;\;\;\;\;\;\;\;\;\;\;\;\;\;\;\;\;(storageCostByQuota)\\
\;\;\;\;\;\;\;\;\;\;\;\;\;\;\;\;\;\;\;\;\;service\_name\& provider\;\;\;\;\;\;\;sum(storage\_\cos t)
\end{array}$}\\
19 & \multicolumn{4}{l}{${\varphi _{storage}}: = storageCost{\bowtie_{provider,location,serviceName}}AvgQoS$}\\
20 & \multicolumn{4}{l}{${\Phi _{network}}: = {\sigma _{provider \in \rho  \wedge location \in \ell  \wedge quot{a_{low}} < {\upsilon _{storage}}}}({\Re _{network}})$ }\\

21 & \multicolumn{4}{l}{//Match appropriate Compute Storage and Network options.}\\
22 & \multicolumn{4}{l}{$\varphi : = {\varphi _{compute}}{\bowtie_{provider,location,locatio{n_{client}}}}{\varphi _{storage}}{\bowtie_{provider,location}}{\Phi _{network}}$ }\\

23 & \multicolumn{4}{l}{$totalCost: = empty\;list$}\\
24 & \multicolumn{4}{l}{$foreach\;\zeta \in \varphi \;do$}\\
25 & $|$ & \multicolumn{3}{l}{$totalCost \cup \left\{ {\zeta \mapsto \sum {{U_r}{P_r}} } \right\}$}\\
26 & \multicolumn{4}{l}{end}\\
27 & \multicolumn{4}{l}{$ranked: = empty\;list$}\\
28 & \multicolumn{4}{l}{$foreach\;\zeta \in totalCost\;do$}\\
29 & $|$ & \multicolumn{3}{l}{$ranked \cup \left\{ {\zeta \mapsto \frac{{\bar \mu {W_{upload}} + \bar D{W_{download}}}}{{\bar \zeta{W_{latency}} + \sum {{W_r}{P_r}} }}} \right\}$ }\\
30 & \multicolumn{4}{l}{end}\\
31 & \multicolumn{4}{l}{return sortOnRankDescending(ranked)}\\

\hline
\end{tabular}
\end{center}
\end{table*}

\subsection{ Algorithm}
\label{subsec:algorithm}
It's more likely that users choose to use a single provider to eliminate costly cross-provider data transfer, but others may have the need to use multiple providers to achieve greater coverage and disaster resilience.

We have abstract our approach in Algorithm 1. Most of the symbols can be found in Table \ref{table:algo_symbols} and \ref{table:relational_algebra_set}, some symbols are defined earlier in Table \ref{table:formula_symbols} and \ref{table:weight_explanation_symbols}. We have separated the relational algebra and set operations into Table \ref{table:relational_algebra_set}, please pay attention to operation G as it has multiple inputs represented by superscript and subscripts. 

Algorithm 1 only depicts one common use case, other scenario exists but can be solved with a simplified version of Algorithm 1 or with small modification/addition. We will explains these situation in the following paragraph.

As shown in Algorithm 1, a user can provide us the following inputs $( \ell , M_{\min } , price_{\max} , \rho , U , W )$. $\ell$ is the set of locations that a user wants to consider, by default we consider all locations. $M_{\min}$ is the minimal memory requirements for the VMs, 0 denotes no memory requirements. $price_{\max}$ is the maximium budget user willing to spend, 0 indicating they are only interested in free services, -1 is used to represent infinity which means there is no budget constrains. $\rho$ is the set of Cloud service providers that a user wants to consider, by default we consider all providers. $U$ represents the the estimated usages of all the resources: $( U_{compute} , U_{storage} , U_{data_{in}} , U_{data_{out}} )$. $U_{compute}$ is the number of instances, $U_{storage}$ is the number of GB of storage will be used. $U_{data_{out}}$ is the amount of outward data transfer in GB from cloud provider to end devices/users. Similarly, $U_{data_{in}}$ represents the amount of inward data transfer. All the previously mentioned usage estimations are all monthly based, but other length can be used such as daily or hourly, as long as all resource are calculated based on the same standard, there should be no effect on the final comparison and ordering. $W$ represents a user's preference, details are explained in section \ref{subsec:cost_benefit_ratio} and \ref{subsec:weight&pairwise_comparison}.

Once options satisfy user requirements have been identified, we calculating price according to different model. There are various pricing models    \cite{weinman2011axiomatic} exist, for example, free, flat-rate, two-part tariffs (like the AWS reserved instance), block-declining (S3 storage), bidding (AWS spot instance). They can mostly be incorporated into our model except the bidding type. One provider often have multiple offers within the same type of services, for example, different kind of instances for the compute service, different storage options, we combine them to get a combinatorial number of choices, we do that for all providers, then calculate the summed cost and rank for each combined option. Not all users need all 3 types of resources, if they specify 0 for a type of resource, it will not be considered. But network service is always needed.

\begin{figure}[!h]
 \centering
 \includegraphics[scale=0.4]{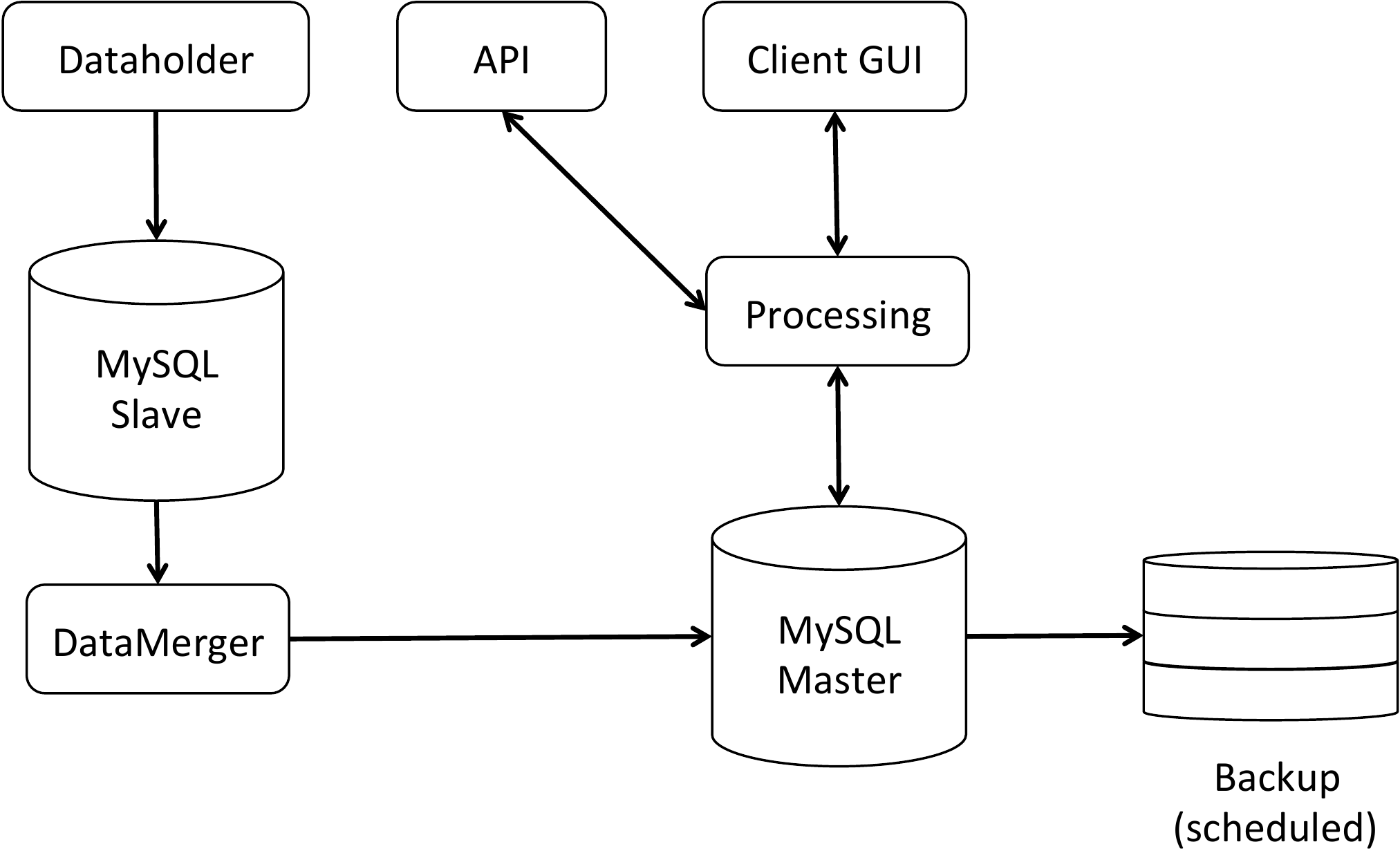}
 \caption{Abstract System Dataflow. This figure is better looking together with figure 3 for better understanding. As we have used several (slave) servers to collect data from different locations. Then we transfer them to a central server for processing and backup, data on this server was also archived and cleared manually every time after we imported the newly collected data into the local offline system for post-processing and cleaning up. We only use the (summarised) average QoS data for real time querying via API and web GUI, as this allows us to provide response faster.}
\label{fig1}
\end{figure}

\begin{figure}[!h]
 \centering
 \includegraphics[scale=0.35]{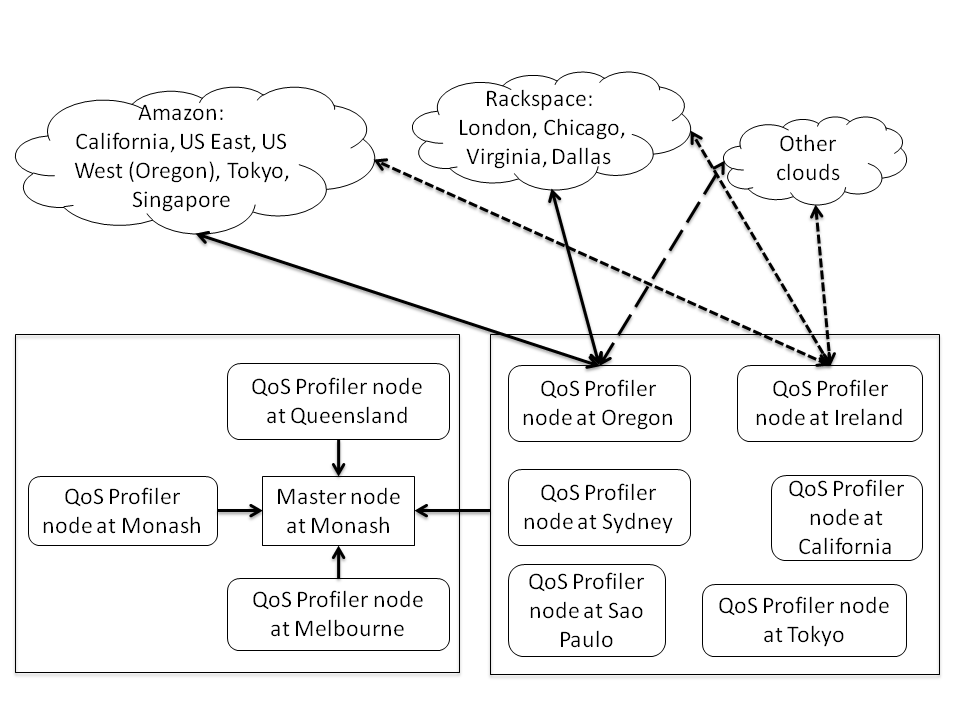}
 \caption{QoS Monitoring Service Network Topology. We have used 2 Clouds namely: Nectar Research Cloud and Amazon Web Service. Since Nectar Cloud is free for researchers, we kept the instances running all the time, hence the decision to put master in Nectar. Because there is a limit of quota in Nectar and Amazon have greater geographical coverage in terms of datacenter locations. We use additional Spot instance from Amazon as slave data crawlers. A QoS Monitoring Node profiles Download Speed, Latency and Upload Speed at each datacenter in various Clouds from different locations.}
\label{fig2}
\end{figure}

\subsection{Implementation}
Fig. \ref{fig1} shows the top level dataflow of the system we implemented. Data is initially collected from web page by profiler nodes, we use the HtmlUnit library \cite{htmlunit}. The whole system consists of multiple agents at geographically dispersed locations to collect and process data, shown in Fig \ref{fig2}. If we look at individual slave node, we can see every node profiles the QoS statistics to various Clouds from each location. Bashed scripts are written to export data from each node. Master node pulls data from its children nodes, access keys are required for this operation. Then the CSV formatted data is imported to the master database, where appropriated merge operation is performed.

\begin{figure}[!h]
 \centering
 \includegraphics[scale=0.4]{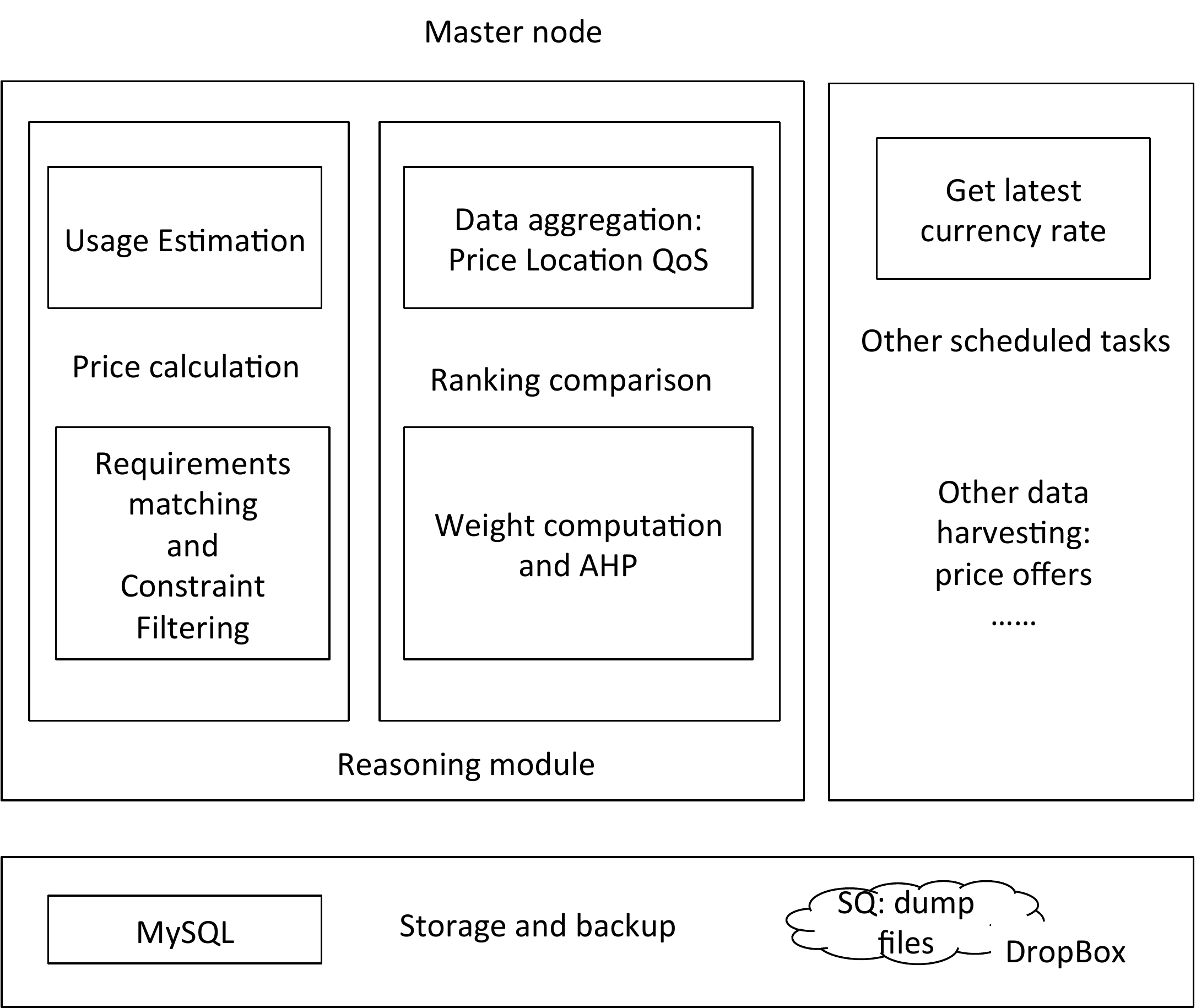}
 \caption{Master Node System Architecture. In the reasoning module main functions and operations are broke down into different blocks. There are some other tasks cannot be strictly categorized into existing modules, those are put into the ``Other Tasks" section, and the very light grey block contains the evolving part of the system so it cannot be considered a stable component of the system. While it’s possible to backup the whole server, it is not necessary at this stage, and the most valuable data is stored in the MySQL database, which can be backuped much easier and cheaper by creating ``SQL dump". This dump file is created daily and simply stored in a Dropbox folder which is free to use and keeps a history of the file stored in it for 30 days, which is sufficient for our case. The presentation layer (UI and API implementation) and monitoring module are omitted to keep the diagram simple.}
\label{fig3}
\end{figure}

Fig. \ref{fig3} shows the overview of our system architecture. We use Dropbox for this prototype implementation to demonstrate the feasibility of our innovation. As long as data is properly backed up in a separate location, other mechanisms can be used.

The price data is collected from providers' websites. The problem with automatic data collection can be solved if providers release more structured data with sufficient metadata description, we have proposed an ontology in previous work \cite{zhang2012ontology}.

Initilally, the QoS data was collected every 2 hours by running the ``speedtest'' service of CloudHarmony. A single run takes more than an hour to finish hence we are collecting it at maximum possible granularity. Later by analyzing the data, we conclude that such high frequency is not necessary, as the average QoS from a particular location to a particular data center most of the time fluctuating between a resealable range. That means the average would be pretty stable. We can use the historical data as a pretty reliable indication. Note that difference between datacenters and various locations are still huge as expected, see Fig. 5. In the future we may allow a combination of real time and off-line values to be used if necessary.

\begin{figure}[!h]
 \centering
 \includegraphics[scale=0.3]{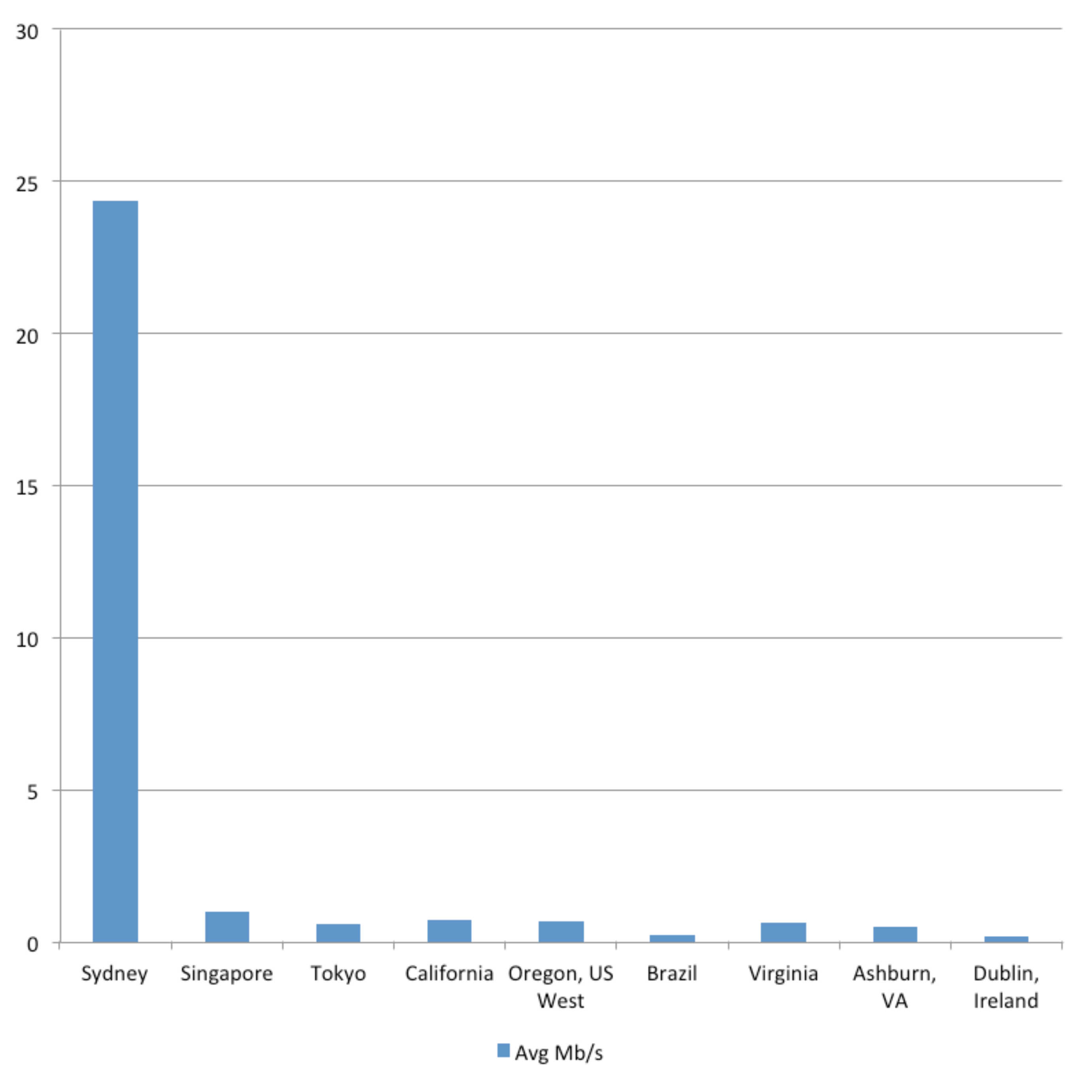}
 \caption{Download speed from Amazon data centers to Melbourne}
\label{fig5}
\end{figure}

\begin{table*}[!htp]
\begin{center}\caption{EXPERIMENT ENVIRONMENTS } \label{table:experiment_env}
\begin{tabular}{|r|l|l|l|l|l|}
\hline
\textbf{Environment}&  \textbf{Description } &  \textbf{Processor Speed }&  \textbf{Memory }&  \textbf{Processor Name }&  \textbf{Role }\\
\hline 1 &  $\;$ MacBook Air Physical machine &    1.4 GHz &  2 GB &  $\;$ Intel Core 2 Duo & Master  \\
\hline 2 & $\begin{array}{l}
    {\rm{Ubuntu}}\;{\rm{12}}{\rm{.04}}{\rm{.3}}\;{\rm{LTS}}\;{\rm{instance}}\;{\rm{in }}\\
    {\rm{a}}\;{\rm{virtualized}}\;{\rm{environment}}
    \end{array}$ &  
    2.4 GHz (1vCPU) 
     &  4 GB& $\begin{array}{l}
    {\rm{AMD}}\;{\rm{Opteron(TM) }}\\
    {\rm{Processor}}\;{\rm{6234}}
    \end{array}$ & Master/Profiler\\
\hline 3 & $\begin{array}{l}
    {\rm{Standard}}\;{\rm{Small}}\;{\rm{(m1}}{\rm{.small) }}\\
    {\rm{Linux/UNIX}}\;{\rm{EC2}}\;{\rm{Spot}}\;{\rm{Instance}}
    \end{array}$ &  
    1.79 GHz (1ECU/vCPU) 
     & 1.7 GB & $\begin{array}{l}
    {\rm{Intel(R)}}\;{\rm{Xeon(R) }}\\
    {\rm{CPU}}\;{\rm{E5 - 2650}}\;
    \end{array}$ & Profiler\\
\hline 4 &  $\begin{array}{l}
    {\rm{Compute}}\;{\rm{Optimized}}\\
    {\rm{(c3}}{\rm{.8xlarge)}}\;{\rm{Linux/UNIX}}\;{\rm{EC2}}\\
    {\rm{Spot}}\;{\rm{Instance }}
    \end{array}$&  
 2.8 GHz (32 vCPU 1081 ECU)
     & 60 GB & $\begin{array}{l}
    {\rm{Intel(R)}}\;{\rm{Xeon(R) }}\\
    {\rm{CPU}}\;{\rm{E5 - 2680 v2}}
    \end{array}$ & Performance Testing \\

\hline
\end{tabular}
\end{center}
\end{table*}

\section{Experiment}\label{experiment}

\subsection{Setup}

We run our system and proposed algorithmic technique across a range of hardware systems to understand the implication of hardware resource configuration (see Table \ref{table:experiment_env}) on the performance of the approach.

To summarize, Environment 1 is the local machine used during the development of the program, which is capable of running the database and other system modules.

Environment 2 is the server from The National eResearch Collaboration Tools and Resources (NeCTAR) cloud \cite{ref40} where the our system can be deployed as a service which is easily accessible over the Internet. It is a virtualized environment, so the CPU speed labeled may not accurately reflect the actual allocation.
NeCTAR's infrastructures are located at at eight different organisations (node sites) around Australia. It operates as one cloud system under the Openstack framework. This makes it having different UI and API compare to AWS. Being a collaborative research cloud, it's only open to affiliated members (i.e. Australian researchers, students from participating university). Although the access is free, there is a limitation of 2 instance per member and a cap on the total resource usage.

Environment 3 is the spot instance type (from Amazon) we used to collect QoS statistics from additional locations, but to cut down the cost; we kept the usage minimal.

Environment 4 is the compute optimized spot instance type we used to test program performance under a powerful CPU, or vertical scalability in short.

\begin{figure}[!h]
 \centering
 \includegraphics[scale=0.3]{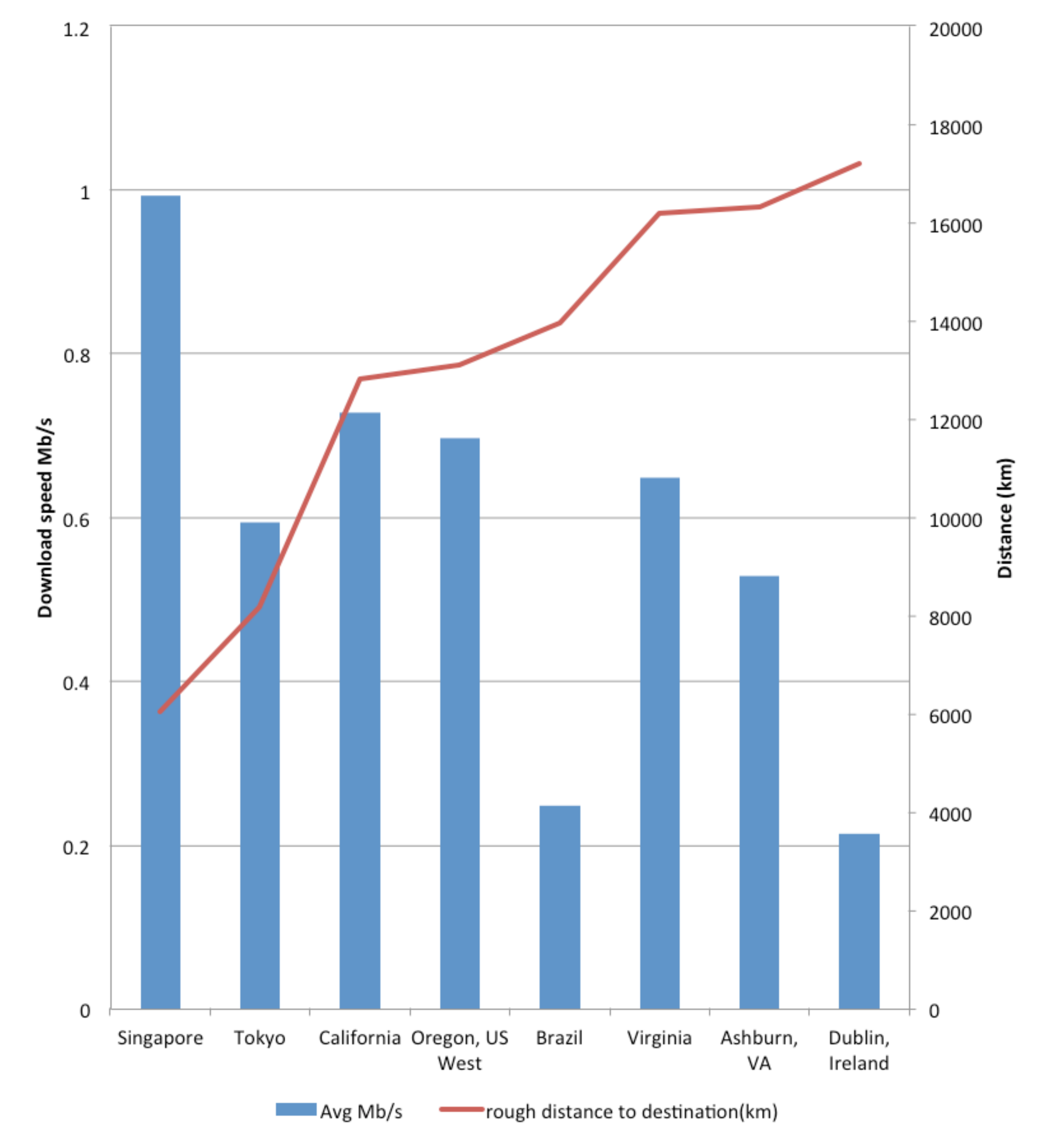}
 \caption{Download speed against distance.}
\label{fig6}
\end{figure}
 
\subsection{Network QoS Data}
Figure \ref{fig5} shows that geographically close data center has (as high as 25 times) better network performance, hence this validates the fact that location is one of the important criteria which should be considered during selection process. Our measurements also indicate that distance is not the only factor that effects the network performance, as shown in Fig. \ref{fig6}, data centers are ordered from closest to furthest from left to right, Tokyo and Brazil clearly perform poorly than expected. Hence, we consider the need for active probing and profiling of network QoS from user's endpoint connection to the cloud data centers. By doing so we get clear picture of data centre's network QoS from the users' device that may be deployed across topologically distributed network locations. Note that we have left out Sydney from Fig. \ref{fig6} on purpose. Fig 5 shows the exponential increase in speed between Sydney and Melbourne compare to overseas locations, while Fig 6 shows the linear relationship between downloading speed and distance among overseas locations. We are aware that while it is generally true that the geographical distance between any pair of servers (or users) on the Internet affects the route trip time (RTT), the bandwidth between them is not necessarily determined by the distance, many other aspects can affect the user end QoS, like the last-mile home-connecting technology, local Internet traffic condition. Our measurements are only providing suggestive base for further optimisation, user's actual experience will vary.

\begin{table}[!h]
\begin{center}\caption{INPUT PARAMETERS } \label{table:input_param}
\begin{tabular}{|l|r|}
\hline
\textbf{Compulsory }&  \textbf{Example Value } \\
\hline Storage(GB/30 Days) & 20 \\
\hline Outbound Data Transfer(GB/30 Days) & 50 \\
\hline Min RAM(GB) & 4 \\
\hline \textbf{Optional } & \textbf{Default Value} \\
\hline Provider Brand & Consider All \\
\hline Display Currency  & AUD \\

\hline Number of Hours to run (per Month) & 720 \\
\hline Number of Instance needed (per Month) & 1 \\
\hline Inbound Data Transfer(GB/30 Days) & 1  \\
\hline Weight of Compute Cost(percentile) & $35\%$ \\
\hline Weight of Storage Cost(percentile) & $25\%$ \\
\hline Weight of Network Cost(percentile) &  $35\%$\\
\hline Weight of Latency(percentile) & $5\%$ \\
\hline Weight of Download Speed(percentile) & $70\%$ \\
\hline Weight of Upload Speed(percentile) & $30\%$ \\
\hline Max RAM(GB) & $100\%$ \\

\hline
\end{tabular}
\end{center}
\end{table}

\subsection{Case Study}
\subsubsection{ Input Parameters}
Table \ref{table:input_param} shows the primary configurable parameters of our algorithm. Everyone's requirements regarding the compulsory parameters usually vary. So we choose a range of values to mimic different selection scenarios. In future work, we may conduct user survey to understand the most concerned factors for different type of users, for example we can exposed all possible constrainable parameters via the API but it may not be necessary (not to mention also slows down the processing) and it will only overwhelm the users who only uses the visual interface. Optional parameters are the one tend to be hard to specify (especially for users with less technical background). Default value column shows what we use when not specified.

\begin{figure*}[!htp]
 \centering
 \includegraphics[scale=0.45]{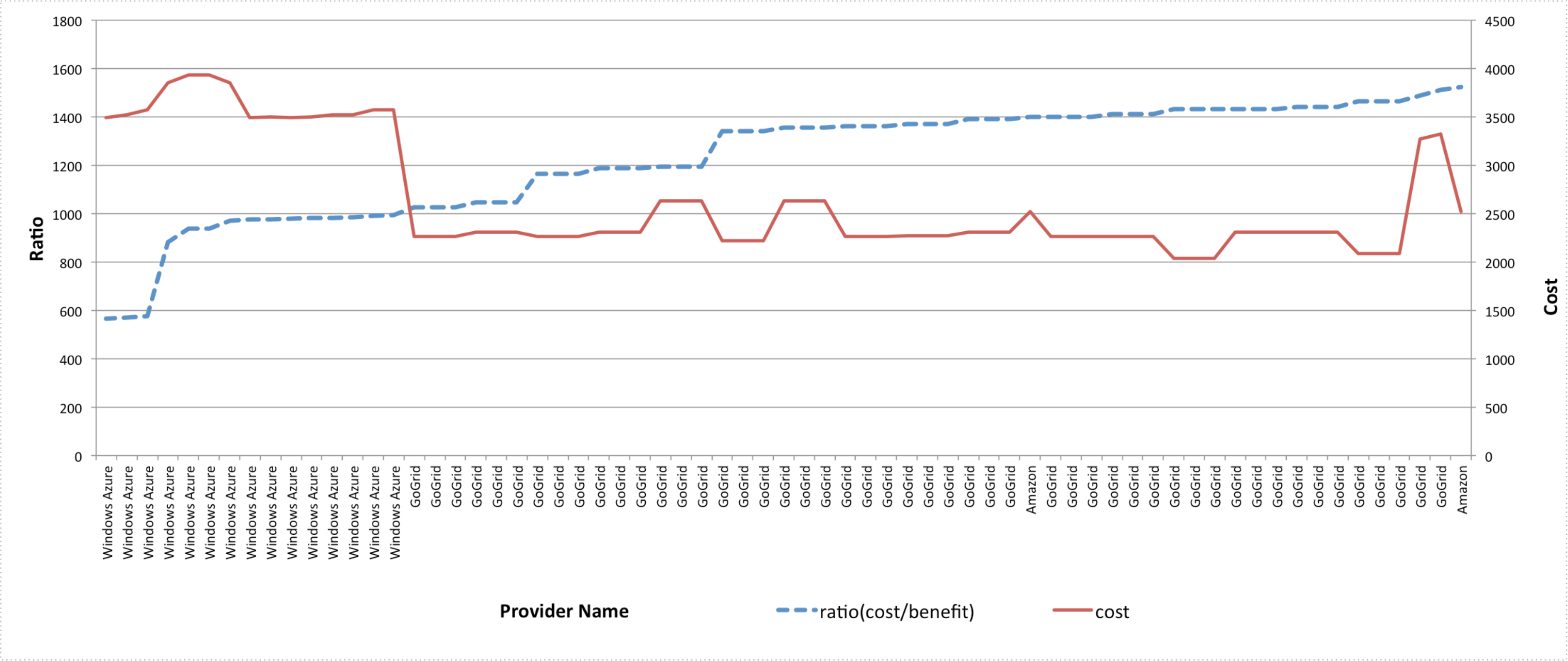}
 \caption{Results in ascending order by (cost / benefit) ratio}
\label{fig7}
\end{figure*}

\subsubsection{Results}

\begin{figure*}[!htp]
 \centering
 \includegraphics[scale=0.5]{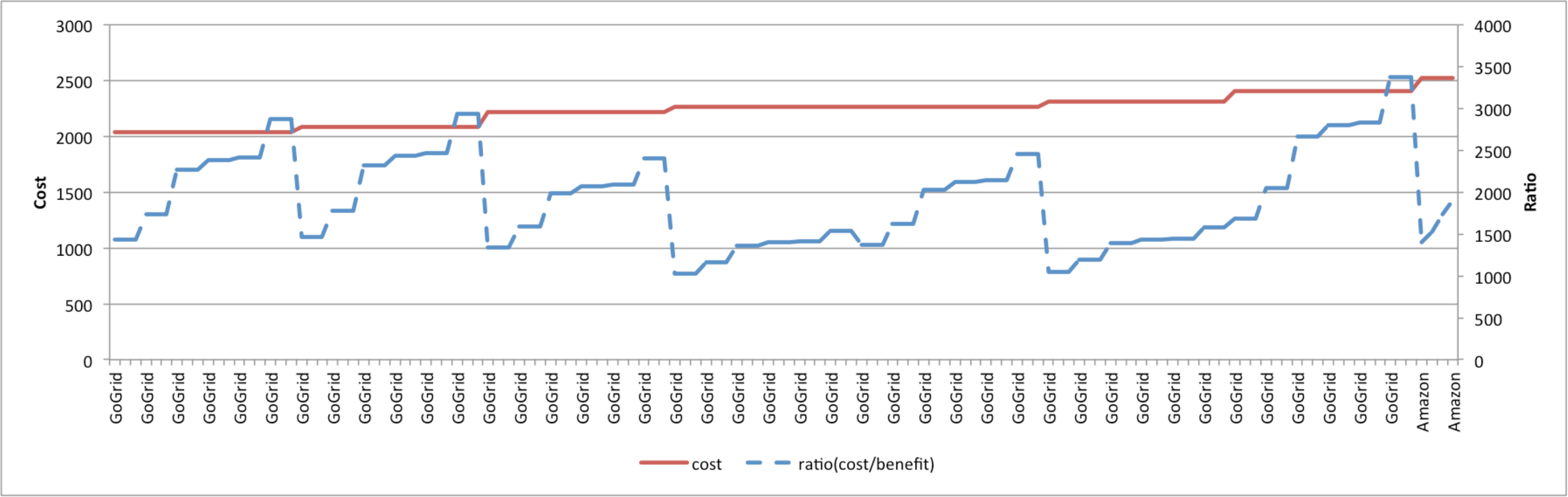}
 \caption{Results in ascending order by cost}
\label{fig8}
\end{figure*}

Figure \ref{fig7} shows the top $5\%$ of the result we get from the inputs in Table \ref{table:input_param}. It is in ascending order of ratio (cost over benefit) as indicated by the dotted (blue) line, because lower cost over higher benefit gives us a smaller ratio which representing a better choice.  If we look at ranking by considering only the cost, as illustrated by the solid (red) line, the GoGrid offers dominate over Windows offerings. If to order results in ascending price order (means network QoS constraints are not considered), shown in Fig. \ref{fig8}, Azure disappears from the top $10\%$ of choices. Similarly, we can see that although the price change is small in solutions, their overall rankings are greatly different (dotted blue line). What this means to users is that while we can save money by ignoring network QoS but then they should be ready for degraded network performance Note that although we tried out best in using real world data, sometimes cloud providers vary their prices as frequent as weekly. However, in future work we intend to implement a price crawler service that will automatically parse the provider's web pages and update our system's database.

\begin{table*}[htbp]
\begin{center}\caption{AVERAGE RUNTIME } \label{table:avg_runtime}
\begin{tabular}{|r|r|r|r|r|r|r|r|}
\hline
\textbf{$\begin{array}{l}
{\rm{Test}}\\
{\rm{Number}}
\end{array}$} &  \textbf{$\begin{array}{l}
{\rm{Storage(GB/}}\\
{\rm{30 Days)}}
\end{array}$ } &  \textbf{$\begin{array}{l}
{\rm{Outbound Data}}\\
{\rm{Transfer(GB/30 Days)}}
\end{array}$ }&  \textbf{$\begin{array}{l}
{\rm{Min}}\\
{\rm{RAM(GB) }}
\end{array}$}&  Row(s) &  Enviroment 1&  Enviroment 2&  Enviroment 4\\
\hline 1 & 20 & 10 & 0 & 3808 & 12.04 & 11.07 & 10.96   \\
\hline 2 & 40 & 15 & 0 & 3808 & 11.913 & 11.59 & 7.81\\
\hline 3 & 10 & 2 & 0 & 3808 & 11.169 & 10.76 & 7.05 \\
\hline 4 & 20 & 2 & 0  & 3808 & 11.744 & 11.15 & 7.57 \\
\hline 5 & 200 & 200 & 0  & 3808 & 11.894 & 11.72 & 7.49 \\
\hline 6 & 200 & 200 & 0  & 3808 & 11.912 & 10.85 & 6.76 \\
\hline 7 & 200 & 200 & 16  & 552 & 9.15 & 7.7 & 4.97 \\
\hline 8 & 200 & 200 & 8  & 1524 & 9.644& 9.69 & 5.53 \\
\hline 9 & 200 & 200 & 4  & 2095 & 10.25 & 8.72 & 5.58 \\
\hline 10 & 20 & 20 & 0  & 3808 & 12.06 & 11.51 & 7.03 \\
\hline \multicolumn{5}{|r|}{Average} & 11.1776 & 10.476 & 7.075\\
\hline
\end{tabular}
\end{center}
\end{table*}

\subsubsection{Performance}

The average run time for our current solution is about 11 seconds, with cache turned on in MySQL, we get up to $9\%$ improvement on the same query. As the constraints become stricter the solution space reduces, as a result processing time decreases (to as low as 4.97 seconds), see Table \ref{table:avg_runtime}.

The performance increase observed when we move from environments 1 to 2 then 4 is resulted from an increase of processing power, hence the idea of ``scale up". There is a limit to the amount of processing power one core can have, but our solution is single threaded at the moment, there is still room for improvement by utilizing all cores (like environment 4).  In the future we will explore the option of configuring MySQL/InnoDB to use multithreads (Default is 4 and maximum is 64 since MySQL 5.1.38). Then we will decide whether we need to ``scale out".

\subsection{Computational Complexity}

We define the upper bound computational complexity of our optimization approach as:
\begin{equation}
O\left( {\left| R \right| \times \left| C \right| \times \left| L \right| + \left( \frac{(|\nu|-1)|\nu|}{2}\right)} \right)
\end{equation}

In cost estimation, we have to calculate prices for $|R|$ resources, $|C|$ providers, and $|L|$ geographical locations. In case a more complex utilization function is given, the computational complexity may increase.

In our current model, we consider $|v|$= 6, see weight calculation in Table \ref{table:weight} on page \pageref{table:weight}. Hence to determine the weights in the benefit-cost ratio evaluation function, 15 pair-wise comparisons have to be made, unless user choose to use the default values. In both cases this part of the complexity factor is a constant which can be omitted.

\section{CONCLUSION AND FUTURE WORK}\label{conclusion}

The cloud has great potential for a large variety of users with diverse needs, but the selection of a the right provider is crucial to this end. Aiming to eliminate potential bottlenecks that limit the ability of general users to take advantage of cloud computing, we present an improved system (which extended out previous work) that further allows user to make multi-criteria selection and comparison on IaaS offers considering QoS. We hope our research will drive even greater adoption of the cloud and boost the expansion of the cloud hosted applications. Furthermore, the system we are proposing will also benefit the Cloud provider, by providing analyses of the market and demand, our system can potentially recommend what price the providers can set their service to.

In the future, we would like to provide smarter decision support by including SLA, legal compliance    \cite{mouratidis2013framework} into consideration. We are also improving the data gathering and updating mechanism. Furthermore, we plan to conduct our experiments on network QoS data collected in real-time rather than based on archived QoS (as done in this paper). This will allow us to analyze performance of the proposed technique under uncertainties such as network congestion and network link failures. There are also other interesting ordinal optimization based techniques\cite{OOZhang2014}, \cite{MOSZhang2014309} worth looking at.
 
\section{Acknowledgement}
Dr. Lizhe Wang's work in this paper is supported by  the Natural Science Foundation of Hebei Province under Gant (No.F2014203093).
Dr. Rajiv Ranjan's work in this paper is funded by Australia India Strategic Research Grant titled "innovative solutions for big data and disaster management applications on clouds (AISRF-08140)".

\bibliographystyle{ieeetr}
\bibliography{lizhe}

\end{document}